\documentclass[aps,prd,10pt,nofootinbib,twocolumn]{revtex4-1}
\usepackage{amsmath,amssymb,amsfonts,dsfont,mathrsfs,amsthm,mathtools}
\usepackage{graphicx}
\usepackage{hyperref}
\usepackage{siunitx}
\usepackage[english]{babel}
\hypersetup{linktocpage,colorlinks=true,urlcolor=blue,linkcolor=blue,citecolor=blue}

\begin{document}

\title{Explicit Lorentz violation in a static and spherically-symmetric spacetime}

\author{Yuri Bonder}
\email{bonder@nucleares.unam.mx}
\author{Christian Peterson}

\affiliation{Instituto de Ciencias Nucleares, Universidad Nacional Aut\'onoma 
de M\'exico\\ Apartado Postal 70-543, Cd.~Mx., 04510, M\'exico}

\begin{abstract}
Lorentz violation is motivated by quantum gravity and it is generically described by nondynamical tensors. In this work a Lorentz violating extension of general relativity is studied where a nondynamical tensor couples to the Weyl tensor. A family of static and spherically symmetric solutions in vacuum is found, confirming that there are consistent solutions with explicit Lorentz violation in dynamical spacetimes. These solutions produce an unconventional dependence of the gravitational redshift, which, in turn, leads to the first bounds on such nondynamical tensor that do not rely on the physics of the early universe. Moreover, the bounds obtained in this work are competitive with respect to limits on similar nondynamical tensors.
\end{abstract}

\maketitle

\section{Introduction}

Local Lorentz invariance lies at the core of general relativity (GR). This principle states that two identical (local) experiments performed in two different inertial frames must yield identical results. It also implies that there are no preferred spacetime directions, which are typically introduced through nondynamical tensors. Being one of the pillars of modern physics, it is important to test local Lorentz invariance empirically. In addition, in most quantum gravity proposals there are compelling arguments that support the idea that local Lorentz invariance is broken (see, e.g., Refs.~\onlinecite{KosteleckySamuel,GambiniPullin}). Thus, some aspects of the quantum nature of gravity could be uncovered by looking for Lorentz violation.

The so-called Standard Model Extension \cite{SME1,SME2,Kostelecky2004} (SME) is a framework that parametrizes all Lorentz violations systematically. The SME action is built using effective field theory \cite{Rob}, therefore, its action contains all terms that can be built with the spacetime metric and the Standard Model (SM) fields that are independent under coordinate and SM gauge transformations, but it includes Lorentz violating terms. The size of each Lorentz violating term is controlled by the SME `coefficients', for which there are many experimental bounds \cite{DataTables}.

Conveniently, the SME can be separated into sectors inherited from conventional physics. This paper studies the gravity sector of the SME \cite{Kostelecky2004}. It is also assumed that gravity is completely described by a four-dimensional pseudo-Riemannian metric, i.e., no additional gravitational degrees of freedom, like torsion, are considered. What is more, the SME has a natural hierarchy for its action terms given by their mass dimension; those terms of the smaller mass dimensions comprise the so-called minimal sector.

The action for the minimal gravity SME sector, with conventional matter, takes the form \cite{Kostelecky2004}
\begin{equation}\label{action}
S=\frac{1}{2 \kappa}\int d^4 x \sqrt{-g}\left( R + k^{abcd}R_{abcd}\right)+ S_{\rm matter},
\end{equation}
where $\kappa$ is GR's coupling constant and Latin indexes are (abstract) spacetime indexes. Also, $g_{ab}$ is the spacetime metric (of signature $-2$) whose determinant is denoted by $g$. Following the conventional practice, indexes are lowered and raised with the metric and its inverse $g^{ab}$ and repeated indexes imply their contraction. ${R_{abc}}^d$ is the Riemann curvature tensor associated with $g_{ab}$, $R_{ab}={R_{acb}}^c$ is the Ricci tensor and $R=g^{ab}R_{ab}$ is the curvature scalar. All the geometric tensors are defined following the conventions of Ref. \onlinecite{Wald}. Moreover, geometrized units where $G=1=c$ are utilized. Finally, $k^{abcd}$ are the SME coefficients, which have the same index symmetries than $R_{abcd}$, and $S_{\rm matter}$ is the matter action. Note that $k^{abcd}$ is dimensionless, but no other constraints are imposed \textit{a priori} on it, in particular, $k^{abcd}$ can have nonvanishing derivatives.

To look for the phenomenological consequences of the action \eqref{action} it is useful to split $k^{abcd}R_{abcd}$ into irreducible pieces
\begin{equation}
k^{abcd}R_{abcd}=-u R + s^{ab}R_{ab} + t^{abcd}W_{abcd},
\end{equation}
where $W_{abcd}$ is the Weyl tensor, that is, the completely traceless part of $R_{abcd}$. In this setup $u$, $s^{ab}$, and $t^{abcd}$ are the SME coefficients, which are traceless. Moreover, the index symmetries inherited from $k^{abcd}$ imply that $s^{ab}$ and $t^{abcd}$ have, respectively, nine and ten independent components.

Clearly, the $u$ term does not produce Lorentz violations; it merely changes the coupling constant. In addition, there are strong constraints on $s^{ab}$ \cite{PhysRevLett.99.241103,PhysRevLett.100.031101,PhysRevD.80.016002,PhysRevD.88.102001,PhysRevLett.112.111103,kostelecky2015constraints,PhysRevD.94.084002,PhysRevLett.117.071102,PhysRevLett.119.201101,PhysRevLett.119.201102,PhysRevD.97.024019,PhysRevD.99.104062,PhysRevD.99.075032} (see also Ref.~\onlinecite{reviewbounds} for a review). However, for reasons that are explained in the next paragraph, the $t^{abcd}$ coefficient remains essentially unconstrained. Also, it has been shown that $u$ and $s^{ab}$ can be ``moved'' to the matter action through metric redefinitions \cite{Bonder2015}. For these reasons, only the $t^{abcd}$ coefficient is considered here.

As is evident from the action \eqref{action}, the metric variation produces a modified Einstein equation with terms containing $k^{abcd}$. Now, the Einstein tensor and the matter contribution (energy-momentum tensor) are divergence free by virtue of the Bianchi identity and the fact that $S_{\rm matter}$ does not contain nondynamical fields \cite{Cristobal1}. Yet, there is no reason for the terms containing $k^{abcd}$ to have a vanishing divergence, as it is required for consistency. This issue motivated the SME community to argue that, when spacetime is dynamical, Lorentz violation must arise spontaneously \cite{Kostelecky2004,PhysRevD.91.065034}. Namely, that $k^{abcd}$ has to be dynamical and its dynamics must select preferred spacetime directions. Many interesting results have been found in the context of spontaneous Lorentz violation \cite{KosteleckyBlumm1,KosteleckyBlumm2,PhysRevD.77.125007,PhysRevD.79.029902,BonderEscobar}. However, the procedure developed to study such violations in curved backgrounds \cite{BaileyKostelecky2006} is rather cumbersome, mainly because the detailed dynamics for the coefficients is unknown. Remarkably, when this procedure is applied to the $t^{abcd}$ term, it yields a trivial relation of the form $0=0$, from which no physical information can be extracted \cite{BaileyKostelecky2006,tpuzz2,AltschulBaileyKostelecky2010,tpuzz3,tpuzz4,tpuzz5}. This fact is known in the SME as the `$t$-puzzle' \cite{BaileyKosteleckyXu2015} and, for some time, people were speculating if $t^{abcd}$ could be unphysical \cite{Bonder2015}, until it was shown that it produces physical effects in cosmology \cite{Gabriel}. Still, the question of what are the effects of $t^{abcd}$ at ``smaller'' scales remained unanswered; tackling this question is the main goal of this paper.

It is also assumed that Lorentz violation, even in a dynamical spacetime, can be explicit (see also Ref.~\onlinecite{Explicit}). This assumption is motivated by the following arguments: first, explicit Lorentz violation does not necessarily brake invariance under diffeomorphisms \cite{Cristobal2}, in contrast to the case where the violation is spontaneous \cite{KosteleckyBlumm2}. Second, other gravitational degrees of freedom that have not been observed (e.g., torsion) can ``absorbe'' the restrictions from the divergence-free condition \cite{Cristobal2}, thus allowing the SME coefficients to be general. And third, with the perturbation scheme employed here, when Lorentz is spontaneously violated the spacetime background must be conformally flat \cite{Gabriel} (i.e., the background metric has to produce a vanishing Weyl tensor). In contrast, when Lorentz is explicitly broken, this restriction is not present as there is no equation of motion for $t^{abcd}$. Therefore, by assuming explicit Lorentz violation it is posible to look for the effects of the SME coefficients in general spacetimes, including those that are static and spherically symmetric.

This paper is organized as follows: in Sec. \ref{EOM} the metric equation of motion is obtained. Then, in Sec. \ref{syms}, the symmetries under consideration are implemented. Next, a comparison with experimental data is provided in Sec. \ref{exp} and the outlook and concluding remarks are presented in Sec. \ref{conc}.

\section{Equation of motion}\label{EOM}

In this section the equation of motion for the action \eqref{action} is derived in the case where $k^{abcd}=t^{abcd}$. In addition, its divergence is obtained, which gives a consistency relation. A stationary variation with respect to the inverse metric generates an Einstein tensor $G_{ab}=R_{ab}-R g_{ab}/2$ and the energy-momentum tensor $T_{ab}=-1/(2\sqrt{-g})\delta S_{\rm matter}/\delta g^{ab}$, for the Einstein-Hilbert and matter parts of the action, respectively. It is then necessary to obtain the variation of the $t^{abcd}$ terms. Note that, whenever $W_{abcd}$ is completely contracted with the (traceless) tensor $t^{abcd}$, it can be interchanged with $R_{abcd}$; this is particularly useful when performing the action variation. The variation of $\sqrt{-g}t^{abcd}R_{abcd}$ has three pieces: one from the volume element $\sqrt{-g}$, one related to the fact that the Riemann tensor is actually of rank $(1,3)$ and the index in the upper position is lowered with a metric, and one from the actual variation of the Riemann tensor. The first and second parts of this variation yield, respectively, $-\sqrt{-g}t^{cdef}R_{cdef} g_{ab}/2$ and $-\sqrt{-g}R_{cde(a}{t^{cde}}_{b)}$, where indexes under parenthesis are symmetrized with a $1/2$ factor.

To calculate the remaining variation, the fact that $\delta {R_{abc}}^d=-g^{de}\nabla_{[a}(\nabla_{b]} \delta g_{ec}+\nabla_{|c|} \delta g_{b]e}-\nabla_{|e|} \delta g_{b]c})$ is used, where square brackets denote antisymmetrization, with a $1/2$, and indexes in between bars are not antisymmetrized. The last expression implies
\begin{equation}
\sqrt{-g}{t^{abc}}_d\delta {R_{abc}}^d=-2\sqrt{-g}t^{abcd}\nabla_{a}\nabla_{c} \delta g_{bd}.
\end{equation}
Now, neglecting boundary terms, as is the case throughout this work, and using the relation of the variation for the metric and the inverse, this equation becomes
\begin{equation}
\sqrt{-g}{t^{abc}}_d\delta {R_{abc}}^d=-2\sqrt{-g}\nabla_{c}\nabla_{d}{{t^{c}}_{(ab)}}^{d} \delta g^{ab}.
\end{equation}

After collecting all terms, it can be seen that the metric equation of motion is
\begin{equation}\label{Einstein Mod}
G_{ab}-\frac{g_{ab} }{2} t^{cdef}R_{cdef} -R_{cde(a}{t^{cde}}_{b)}-2\nabla_{c}\nabla_{d}{{t^{c}}_{(ab)}}^{d}=\kappa T_{ab}.
\end{equation}
Observe that it is possible to replace $R_{abcd}$ by $W_{abcd}$ in the second term on the left-hand side, but the third term does depend on $R_{ab}$. Also, due to index symmetries, the two derivatives in the last term act symmetrically on $t^{abcd}$.

If $S_{\rm matter}$ is invariant under diffeomorphisms, i.e., $S_{\rm matter}$ does not depend on nondynamical fields \cite{Cristobal1}, it can be shown that the energy-momentum tensor is divergence-free \cite[page 456]{Wald}. Thus, the divergence of Eq. \eqref{Einstein Mod} is
\begin{eqnarray}
0&=&\frac{1}{2} \nabla^a\left( t^{bcde}R_{bcde}\right) +\nabla_b\left({R_{cde}}^{(a}t^{|cde|b)}\right)\nonumber\\&&
+2\nabla_b\nabla_{c}\nabla_{d}t^{c(ab)d},\label{Bianchi1}
\end{eqnarray}
where the Bianchi identity is used. Interestingly, using the index symmetries of $t^{abcd}$, the fact that $t^{[abc]d}=0$, and that two antisymmetric derivatives can be replaced by the Riemann tensor, the last term can be brought to the form
\begin{eqnarray}
2\nabla_b\nabla_{c}\nabla_{d}t^{c(ab)d}&=& \nabla_b\nabla_{[c}\nabla_{d]}t^{cdba} + 2\nabla_{[b}\nabla_{c]}\nabla_{d}t^{bcda}\nonumber\\
&=&\nabla_b\left({R_{cde}}^{[a} t^{|cde|b]}\right)+ {R_{cde}}^a \nabla_{b} t^{cdeb}.\nonumber\\&&\label{DDt}
\end{eqnarray}
Then, the consistency condition becomes
\begin{equation}\label{Bianchi}
0=\frac{1}{2} \nabla_a\left(R_{bcde} t^{bcde}\right) +\nabla_b\left(R_{cdea}t^{cdeb}\right)+ R_{cdea} \nabla_{b} t^{cdeb},
\end{equation}
where, again, the Riemann tensor in the first term can be replaced by $W_{abcd}$.

Notice that Eq. \eqref{Bianchi} has no additional information than Eq.~\eqref{Einstein Mod}. However, it is extremely useful as it only contains first derivatives. Nevertheless, finding analytical solutions of the theory is nontrivial and one must introduce symmetries to achieve this goal. This is done in the next section.

\section{Static and spherically-symmetric solutions in vacuum}\label{syms}

In this section a solution for Eq. \eqref{Einstein Mod} is obtained in vacuum, i.e., outside gravitational sources. It is also assumed that the solution is static and spherically symmetric. Static means that there is a timelike Killing field that is hypersurface orthogonal \cite{Wald}, and it is convenient to use the parameter along its integral curves as a time coordinate $t$. In terms of this coordinate, the assumption of hypersurface orthogonality implies that all fields are unchanged if $t$ goes to $-t$. On the other hand, spherical symmetry, loosely speaking, means that there are three spacelike Killing vectors satisfying the $\mathfrak{so}(3)$ algebra.

It is widely known \cite[page 121]{Wald} that in the coordinates adapted to this symmetries, $t,r,\theta,\phi$, the metric takes the form
\begin{equation}\label{Metrica}
  \text{d}s^2 = -f\text{d}t^2 + h\text{d}r^2 + r^2(\text{d}\theta^2 +\sin^2\theta \text{d} \phi^2),
\end{equation}
where $f=f(r)$ and $h=h(r)$ are arbitrary functions of the `radial' coordinate $r$. The symmetries are imposed on $t^{abcd}$ by requiring its Lie derivative along all Killing fields to vanish \cite{Frolov}. The nonzero components of the most general $t^{abcd}$ that is static and spherically-symmetric are
\begin{align}
 t^{trtr}  &=  T,           \\
 t^{t \theta t \theta}  & = t^{t \phi t \phi}\sin^2\theta=  \frac{-h T}{2r^2}, \\
 t^{r \theta r \theta}  &= t^{r \phi r \phi} \sin^2\theta=  \frac{f T}{2r^2},\\
  t^{\theta\phi\theta\phi}  &=   \frac{-f h T}{r^4}  \csc^2\theta .
\end{align}
where $T=T(r)$ is another arbitrary function. Notably, the fact that $t^{abcd}$ has only one independent component can be understood from making further decompositions that are analogous to those done to the Weyl tensor \cite{Hall,Bonder2008}. Even though the spacetime region under consideration has no matter fields, for mathematical purposes, $t^{abcd}$ acts as matter in an Einstein equation. Therefore, Birkhoff's theorem, which would imply that spacetime is Schwarzschild, does not apply, and, generically, there are no reasons for $f h$ to be one \cite{Marcelo,Jacobson}.

The equations are solved using a perturbative scheme, which is justified by the fact that, in any reference frame that is relevant for testing local Lorentz invariance, $t^{abcd}$ should be small since, to date, there is no empirical evidence of its violation. Thus, the metric is separated into a background metric and a perturbation that is of the same order than $t^{abcd}$. Now, to keep track of the perturbative orders, a parameter $\epsilon$ is introduced. Assuming a Schwarzschild background, to first order in $\epsilon$, the metric and the $t^{abcd}$ components become
\begin{eqnarray}
f &  =& \left(1-\frac{2M}{r}\right)\left(1+\epsilon\delta f\right),\\
h &= &\frac{1+\epsilon \delta h}{1-\frac{2M}{r}},\\
T & =&\epsilon \delta T,
\end{eqnarray} 
where $M$ is a positive parameter that represents the mass of the star or planet sourcing gravity, and $\delta f$, $\delta h$, and $\delta T$ are functions of $r$ that are determined below.

Equation \eqref{Bianchi} is linear in $t^{abcd}$, therefore, all the geometrical objects in this equation can be replaced by the corresponding background tensors. Moreover, as it is expected from the symmetries, Eq.~\eqref{Bianchi} has only one nontrivial component: the $r$ component. This equation then becomes
\begin{equation}
\frac{12 M \left(2 r \delta T' - 3 \delta T\right)}{r^4}=0,
\end{equation}
where the prime denotes a derivative with respect to $r$. Since $r>0$, this equation is readily solved yielding $\delta T \propto r^{3/2}$. Absorbing the undetermined factor in $\epsilon$, it is possible to take
\begin{equation}\label{solt}
\delta T = \left(\frac{r}{M}\right)^{3/2}.
\end{equation}
Recall that the fact that the SME coefficient takes a particular form is a direct consequence of the divergence-free condition, which, in turn, is due to the fact that the Lorentz violation is explicit.

The solution \eqref{solt} is inserted into Eq.~\eqref{Einstein Mod}, which, with the symmetries under consideration, contains three independent nontrivial equations: $tt$, $rr$, and $\theta\theta$. Surprisingly, the $tt$ component of Eq.~\eqref{Einstein Mod} only involves $\delta h$:
\begin{equation}\label{EqDeltaH}
0=\frac{r-2 M}{r^3} \left[(r-2M) \delta h'+ \delta h+\frac{9}{2}\sqrt{\frac{r}{M}}\left( \frac{5r}{M}-12\right) \right].
   \end{equation}
Since attention is set on the $r>2 M$ spacetime region, which describes the exterior of the source, the terms inside the brackets in Eq.~\eqref{EqDeltaH} need to vanish. The solution is
\begin{equation}
\delta h =\frac{-9 \left(r/M\right)^{3/2} (r-4 M)+c_1 M}{r-2 M},
\end{equation}
with $c_1$ a dimensionless integration constant. Using this last equation, the $rr$ component of Eq.~\eqref{Einstein Mod} takes the form
\begin{equation}
\delta f'=\frac{c_1 M^2-9 \sqrt{r/M} \left(2 M^2-7 M r+2 r^2\right)}{M (r-2 M)^2}.
\end{equation}
This equation can be integrated, leading to
\begin{equation}
\delta f=\frac{-6 (r/M)^{3/2} (2 r-M)-c_1 M}{r-2 M}+c_2,
\end{equation}
where $c_2$ is another integration constant. Importantly, it is possible to verify that the other independent equation, namely, the $\theta\theta$ component of Eq.~\eqref{Einstein Mod}, is automatically satisfied, as expected. Note that, by first solving Eq.~\eqref{Bianchi}, no second order equation has to be solved.

Importantly, for a given $M$, all the vacuum static and spherically symmetric solutions of the theory described by the action \eqref{action} have been found, in the perturbative scheme under consideration, and are completely characterized by three parameters: $\epsilon$, $c_1$, and $c_2$.

It is tempting to claim that the solutions that are found are problematic because $\delta T$, $\delta h$, and $\delta f$ diverge as $r\to \infty$. What is more, the curvature scalar $R$ also diverges when $r \to 2M$. However, in these spacetime regions, the functions $\delta T$, $\delta h$, and $\delta f$ become arbitrarily large, which brakes the perturbative scheme under consideration. Therefore, to study the asymptotic and singular spacetime structure one needs to solve the exact theory. Nevertheless, if one eventually finds that there are no asymptotically flat solutions, it would imply that the theory cannot properly describe isolated bodies and it would shed light into the aforementioned $t$-puzzle since, in Ref.~\onlinecite{BaileyKostelecky2006}, asymptotic flatness is imposed \textit{a priori}.

In the next section, the phenomenological implications of the theory in the vicinity of the Earth are analyzed, leading to bounds on $t^{abcd}$.

\section{Experimental constraints}\label{exp}

Typically, to constrain a static and spherically symmetric geometry, it is possible to perform a post-Newtonian expansion and compare with the constraints on the PPN parameters \cite{Will}. However, in the case at hand, there is no direct translation to the PPN parameters because the theory has no Newtonian limit. This can be understood by noting that, for a generic $t^{abcd}$ and when $T_{ab}=0$, Minkowski is no solution of Eq.~\eqref{Einstein Mod}. Fortunately, the presence of Killing vector fields allows one to easily calculate the geodesics \cite{Wald,Hartle} (see Ref.~\onlinecite{TKilling} for an extension of this method), which represent the trajectories of light and material probes.

In fact, the existence of a hypersurface-orthogonal timelike Killing vector field $ (\partial/\partial t)^a$ allows one to follow the textbook calculation to get the frequency shift of light, conventionally known as redshift (see, e.g., Ref.~\onlinecite{Wald}, pages 136-137). This calculation becomes simpler for observers whose velocities are proportional to $ (\partial/\partial t)^a$ and which are connected by a radial null geodesic (i.e., they have the same angular coordinates). If the radial coordinates for the emitter and receptor are, respectively, $r_1$ and $r_2$, then
\begin{equation}
\frac{\omega_2}{\omega_1}=\sqrt{\frac{g_{tt}(r_1)}{g_{tt}(r_2)}}=\sqrt{\frac{1-\frac{2M}{r_1}}{1-\frac{2M}{r_2}}}\left\{1+\frac{\epsilon}{2} [\delta f(r_1)-\delta f(r_2)]\right\},
\end{equation}
where $\omega_1$ and $\omega_2$ are the frequencies of light as measured by the emitter and the receptor, respectively. Recall that this result can be reinterpreted as an $r$-dependent rate of time.

Note that the square root in the last identity coincides with the GR effect, $(\omega_2/\omega_1)_{\rm GR}$, and that $c_2$ cancels out. Now, for the relevant experimental setups $M\ll r_1$ and $M\ll r_2$, thus, to first order in $M/r$, $\Delta \omega=\omega_2-\omega_1$ satisfies
\begin{equation}\label{redshiftGR}
\left(\frac{\Delta\omega}{\omega_1}\right)_{\rm GR}=\frac{M}{r_2}-\frac{M}{r_1},
\end{equation}
which coincides with the difference in the Newtonian potentials. In addition, the $\mathcal{O}(\epsilon)$ factor becomes
\begin{equation}\label{OurAlpha}
\frac{\epsilon}{2}\left[\delta f(r_1)-\delta f(r_2)\right]=\frac{6\epsilon}{M^{3/2}} \left(r_2^{3/2} - r_1^{3/2}\right),
\end{equation}
where it is further assumed that $|c_1| < M/r_1$ and $|c_1 |< M/r_2$ to neglect the terms containing $c_1$.

The best constraint to the gravitational redshift has been obtained with the GREAT experiment \cite{GREAT} which uses two satellites of the European satellite navigation system, known as Galileo, that were (accidentally) delivered on elliptic orbits. These orbits produce a modulation on the gravitational redshift that can be measured using the onboard atomic clocks. To report the experimental limit, a $1+\alpha$ factor is inserted in the right-hand side of Eq.~\eqref{redshiftGR}, where $\alpha$ parameterizes deviations from GR. The main result of this experiment is that $|\alpha|<2\times 10^{-5}$, with a $1\sigma$ statistical significance.

In the theory under consideration, the term given in Eq.~\eqref{OurAlpha} plays the role of $\alpha$, and thus, its absolute value must be smaller than $2\times 10^{-5}$. It turns out that, for both satellites, the semi-major axis and the eccentricity are, respectively, $ 2.8\times 10^7 {\rm \ m}$ and $0.162$ \cite{DatosGalileo}. With this data, the limit on $\alpha$ can be translated to $|\epsilon|<3 \times 10^{-19}$.

Given that the nonvanishing components of $t^{abcd}$ are independent of $c_1$ and $c_2$, it is enough to have a constraint on $\epsilon$ to place bounds on these components. The limit on $\epsilon$ implies that, at the Earth's surface, $|T|=|\epsilon| (r/M)^{3/2}$ must be smaller than $ 1.7 \times 10^{-5}$, therefore
\begin{align*}
\left| t^{trtr}\right|  &< 1.7 \times 10^{-5},           \\
\left| t^{t \theta t \theta}\right|,\ \left|t^{t \phi t \phi}\right|,\ \left|t^{r \theta r \theta}\right|,\ \left|t^{r \phi r \phi}\right| &< 2.1 \times 10^{-19}\  {\rm \ m}^{-2}, \\
\left|  t^{\theta\phi\theta\phi}\right|  &< 1.0  \times 10^{-32}\  {\rm \ m}^{-4},
\end{align*}
where the length units appear to compensate for the fact that the covectors associated with the angular coordinates, which are used to extract the corresponding $t^{abcd}$ components, are dimensionless. Also, for the estimation of these bounds, the trigonometric functions are taken to be of order $1$.

To translate these bounds to a Sun-centered orthonormal frame $(T,X,Y,Z)$, following the SME convention \cite{DataTables}, it is necessary to first change coordinates to isotropic Schwarzschild coordinates \cite{Buchdahl} (the $\epsilon$ effects are neglected because $t^{abcd}$ is order $\epsilon$). Then the spatial spherical coordinates must be made Cartesian, and finally a translation of the origin to the center of the Sun must be performed, in which the effects of curvature and the relative velocity of the Earth with respect to the Sun are neglected. After all these steps are performed, the results become limits on linear combinations of the $t^{abcd}$ components. However, to report the results, only a single nonvanishing component is considered at a time. The best bounds obtained for every component are summarized in Table \ref{table}, where identities between the components are used (see, e.g., Ref.~\cite{Bonder2008}). For the sake of comparison, the limits on the $s^{ab}$ in the same reference frame range from $10^{-11}$ to $10^{-8}$, indicating that the bounds obtained here are competitive.

\begin{table}[h]
  \begin{center}
    \caption{Limits on $t^{abcd}$ in the conventional SME Sun-centered frame.}
    \label{table}
\begin{tabular}{ |c | c| }
\hline
Components & Bound\\[1ex]
\hline 
&\\
  $\left|t^{TXTX}\right|$,$\left|t^{YZYZ}\right|$ &  $<4.4\times 10^{-13}$ \\[1ex]
  $\left|t^{TXTY}\right|$, $\left|t^{TXTZ}\right|$,$\left|t^{YZXZ}\right|$,$\left|t^{YZXY}\right|$ & $<5.0\times 10^{-9}$ \\[1ex]
  $\left|t^{TYTY}\right|$, $\left|t^{TZTZ}\right|$,$\left|t^{XZXZ}\right|$& $<2.7\times 10^{-4}$ \\[1ex]
  $\left|t^{TYTZ}\right|$,$\left|t^{XYXZ}\right|$ &  $<1.5\times 10^{-5}$ \\[1ex]
  $\left|t^{XYXY}\right|$ & $<7.7\times 10^{-5}$ \\
\hline
 \end{tabular}
  \end{center}
\end{table}

\section{Conclusions}\label{conc}

In this work explicit Lorentz violation is studied in the gravity sector. In particular, the effects of the so-called $t$ term are considered in vacuum and for static and spherically symmetric cases, which is a good approximation to describe the exterior of a star or a planet. Remarkably, using a simple perturbative scheme and the Bianchi identity, all the solutions can be found analytically, showing that it is indeed possible to find consistent solutions in a theory with explicit Lorentz violation on a dynamical spacetime. In addition, these solutions give further evidence that the $t$ term does produce physical effects: it modifies the metric dependence on the radial coordinate.

The phenomenological implications of the theory are studied to obtain quantitative limits on the SME coefficients. As expected, the theory produces an unusual dependence of the gravitational redshift on the radial coordinate. This is enough to place bounds on the $t$ coefficient, which are the first bounds placed on this coefficient without appealing to a concrete history of the early universe. Moreover, these bounds, when translated into the conventional SME frame, are competitive with other limits found in the gravity sector.

The empirical bounds obtained here arise from observations in the Earth's gravitational environment. However, the fact that $t$ depends on the mass of the source suggests that it may be more natural to use the Sun as the source, mainly because the Sun-centered frame is the conventional reference frame in the SME. In addition, experiments where the Sun is the gravitational source may lead to better bounds on $t$. Perhaps the most relevant example of these tests is the Shapiro time delay \cite{Shapiro1,Shapiro2}, which yields the most stringent bounds among the `classical' tests of general relativity in the parametrized post-Newtonian framework.

Finally, note that the analysis presented here is insensitive to the $t$ components that have only one time index. This is, of course, a direct consequence of assuming staticity. Therefore, to look for the effects of the remaining $t$ components using planets or stars as the source, it is necessary to include their rotation, which, for the sources at hand, is very small. This indicates that the components with a single time index are not going to be well constrained with extensions of this work.

\begin{acknowledgments}
This research was funded by UNAM-DGAPA-PAPIIT Grant No. IA101818 and CONACyT through the graduate school scholarships.
 \end{acknowledgments}

\bibliography{References}

\begin{thebibliography}{53}%
\makeatletter
\providecommand \@ifxundefined [1]{%
 \@ifx{#1\undefined}
}%
\providecommand \@ifnum [1]{%
 \ifnum #1\expandafter \@firstoftwo
 \else \expandafter \@secondoftwo
 \fi
}%
\providecommand \@ifx [1]{%
 \ifx #1\expandafter \@firstoftwo
 \else \expandafter \@secondoftwo
 \fi
}%
\providecommand \natexlab [1]{#1}%
\providecommand \enquote  [1]{``#1''}%
\providecommand \bibnamefont  [1]{#1}%
\providecommand \bibfnamefont [1]{#1}%
\providecommand \citenamefont [1]{#1}%
\providecommand \href@noop [0]{\@secondoftwo}%
\providecommand \href [0]{\begingroup \@sanitize@url \@href}%
\providecommand \@href[1]{\@@startlink{#1}\@@href}%
\providecommand \@@href[1]{\endgroup#1\@@endlink}%
\providecommand \@sanitize@url [0]{\catcode `\\12\catcode `\$12\catcode
  `\&12\catcode `\#12\catcode `\^12\catcode `\_12\catcode `\%12\relax}%
\providecommand \@@startlink[1]{}%
\providecommand \@@endlink[0]{}%
\providecommand \url  [0]{\begingroup\@sanitize@url \@url }%
\providecommand \@url [1]{\endgroup\@href {#1}{\urlprefix }}%
\providecommand \urlprefix  [0]{URL }%
\providecommand \Eprint [0]{\href }%
\providecommand \doibase [0]{http://dx.doi.org/}%
\providecommand \selectlanguage [0]{\@gobble}%
\providecommand \bibinfo  [0]{\@secondoftwo}%
\providecommand \bibfield  [0]{\@secondoftwo}%
\providecommand \translation [1]{[#1]}%
\providecommand \BibitemOpen [0]{}%
\providecommand \bibitemStop [0]{}%
\providecommand \bibitemNoStop [0]{.\EOS\space}%
\providecommand \EOS [0]{\spacefactor3000\relax}%
\providecommand \BibitemShut  [1]{\csname bibitem#1\endcsname}%
\let\auto@bib@innerbib\@empty
%</preamble>
\bibitem [{\citenamefont {Kosteleck\'y}\ and\ \citenamefont
  {Samuel}(1989)}]{KosteleckySamuel}%
  \BibitemOpen
  \bibfield  {author} {\bibinfo {author} {\bibfnamefont {V.~A.}\ \bibnamefont
  {Kosteleck\'y}}\ and\ \bibinfo {author} {\bibfnamefont {S.}~\bibnamefont
  {Samuel}},\ }\href {\doibase 10.1103/PhysRevLett.63.224} {\bibfield
  {journal} {\bibinfo  {journal} {Phys. Rev. Lett.}\ }\textbf {\bibinfo
  {volume} {63}},\ \bibinfo {pages} {224} (\bibinfo {year} {1989})}\BibitemShut
  {NoStop}%
\bibitem [{\citenamefont {Gambini}\ and\ \citenamefont
  {Pullin}(1999)}]{GambiniPullin}%
  \BibitemOpen
  \bibfield  {author} {\bibinfo {author} {\bibfnamefont {R.}~\bibnamefont
  {Gambini}}\ and\ \bibinfo {author} {\bibfnamefont {J.}~\bibnamefont
  {Pullin}},\ }\href {\doibase 10.1103/PhysRevD.59.124021} {\bibfield
  {journal} {\bibinfo  {journal} {Phys. Rev. D}\ }\textbf {\bibinfo {volume}
  {59}},\ \bibinfo {pages} {124021} (\bibinfo {year} {1999})}\BibitemShut
  {NoStop}%
\bibitem [{\citenamefont {Colladay}\ and\ \citenamefont
  {Kosteleck\'y}(1997)}]{SME1}%
  \BibitemOpen
  \bibfield  {author} {\bibinfo {author} {\bibfnamefont {D.}~\bibnamefont
  {Colladay}}\ and\ \bibinfo {author} {\bibfnamefont {V.~A.}\ \bibnamefont
  {Kosteleck\'y}},\ }\href {\doibase 10.1103/PhysRevD.55.6760} {\bibfield
  {journal} {\bibinfo  {journal} {Phys. Rev. D}\ }\textbf {\bibinfo {volume}
  {55}},\ \bibinfo {pages} {6760} (\bibinfo {year} {1997})}\BibitemShut
  {NoStop}%
\bibitem [{\citenamefont {Colladay}\ and\ \citenamefont
  {Kosteleck\'y}(1998)}]{SME2}%
  \BibitemOpen
  \bibfield  {author} {\bibinfo {author} {\bibfnamefont {D.}~\bibnamefont
  {Colladay}}\ and\ \bibinfo {author} {\bibfnamefont {V.~A.}\ \bibnamefont
  {Kosteleck\'y}},\ }\href {\doibase 10.1103/PhysRevD.58.116002} {\bibfield
  {journal} {\bibinfo  {journal} {Phys. Rev. D}\ }\textbf {\bibinfo {volume}
  {58}},\ \bibinfo {pages} {116002} (\bibinfo {year} {1998})}\BibitemShut
  {NoStop}%
\bibitem [{\citenamefont {Kosteleck\'y}(2004)}]{Kostelecky2004}%
  \BibitemOpen
  \bibfield  {author} {\bibinfo {author} {\bibfnamefont {V.~A.}\ \bibnamefont
  {Kosteleck\'y}},\ }\href {\doibase 10.1103/PhysRevD.69.105009} {\bibfield
  {journal} {\bibinfo  {journal} {Phys. Rev. D}\ }\textbf {\bibinfo {volume}
  {69}},\ \bibinfo {pages} {105009} (\bibinfo {year} {2004})}\BibitemShut
  {NoStop}%
\bibitem [{\citenamefont {Kosteleck\'y}\ and\ \citenamefont
  {Potting}(1995)}]{Rob}%
  \BibitemOpen
  \bibfield  {author} {\bibinfo {author} {\bibfnamefont {V.~A.}\ \bibnamefont
  {Kosteleck\'y}}\ and\ \bibinfo {author} {\bibfnamefont {R.}~\bibnamefont
  {Potting}},\ }\href {\doibase 10.1103/PhysRevD.51.3923} {\bibfield  {journal}
  {\bibinfo  {journal} {Phys. Rev. D}\ }\textbf {\bibinfo {volume} {51}},\
  \bibinfo {pages} {3923} (\bibinfo {year} {1995})}\BibitemShut {NoStop}%
\bibitem [{\citenamefont {Kosteleck\'y}\ and\ \citenamefont
  {Russell}(2011)}]{DataTables}%
  \BibitemOpen
  \bibfield  {author} {\bibinfo {author} {\bibfnamefont {V.~A.}\ \bibnamefont
  {Kosteleck\'y}}\ and\ \bibinfo {author} {\bibfnamefont {N.}~\bibnamefont
  {Russell}},\ }\href {\doibase 10.1103/RevModPhys.83.11} {\bibfield  {journal}
  {\bibinfo  {journal} {Rev. Mod. Phys.}\ }\textbf {\bibinfo {volume} {83}},\
  \bibinfo {pages} {11} (\bibinfo {year} {2011})},\ \bibinfo {note}
  {arxiv:0801.0287v13}\BibitemShut {NoStop}%
\bibitem [{\citenamefont {Wald}(1984)}]{Wald}%
  \BibitemOpen
  \bibfield  {author} {\bibinfo {author} {\bibfnamefont {R.~M.}\ \bibnamefont
  {Wald}},\ }\href@noop {} {\emph {\bibinfo {title} {{General Relativity}}}}\
  (\bibinfo  {publisher} {Chicago University Press},\ \bibinfo {year}
  {1984})\BibitemShut {NoStop}%
\bibitem [{\citenamefont {Battat}\ \emph {et~al.}(2007)\citenamefont {Battat},
  \citenamefont {Chandler},\ and\ \citenamefont
  {Stubbs}}]{PhysRevLett.99.241103}%
  \BibitemOpen
  \bibfield  {author} {\bibinfo {author} {\bibfnamefont {J.~B.~R.}\
  \bibnamefont {Battat}}, \bibinfo {author} {\bibfnamefont {J.~F.}\
  \bibnamefont {Chandler}}, \ and\ \bibinfo {author} {\bibfnamefont {C.~W.}\
  \bibnamefont {Stubbs}},\ }\href {\doibase 10.1103/PhysRevLett.99.241103}
  {\bibfield  {journal} {\bibinfo  {journal} {Phys. Rev. Lett.}\ }\textbf
  {\bibinfo {volume} {99}},\ \bibinfo {pages} {241103} (\bibinfo {year}
  {2007})}\BibitemShut {NoStop}%
\bibitem [{\citenamefont {M\"uller}\ \emph {et~al.}(2008)\citenamefont
  {M\"uller} \emph {et~al.}}]{PhysRevLett.100.031101}%
  \BibitemOpen
  \bibfield  {author} {\bibinfo {author} {\bibfnamefont {H.}~\bibnamefont
  {M\"uller}} \emph {et~al.},\ }\href {\doibase 10.1103/PhysRevLett.100.031101}
  {\bibfield  {journal} {\bibinfo  {journal} {Phys. Rev. Lett.}\ }\textbf
  {\bibinfo {volume} {100}},\ \bibinfo {pages} {031101} (\bibinfo {year}
  {2008})}\BibitemShut {NoStop}%
\bibitem [{\citenamefont {Chung}\ \emph {et~al.}(2009)\citenamefont {Chung}
  \emph {et~al.}}]{PhysRevD.80.016002}%
  \BibitemOpen
  \bibfield  {author} {\bibinfo {author} {\bibfnamefont {K.-Y.}\ \bibnamefont
  {Chung}} \emph {et~al.},\ }\href {\doibase 10.1103/PhysRevD.80.016002}
  {\bibfield  {journal} {\bibinfo  {journal} {Phys. Rev. D}\ }\textbf {\bibinfo
  {volume} {80}},\ \bibinfo {pages} {016002} (\bibinfo {year}
  {2009})}\BibitemShut {NoStop}%
\bibitem [{\citenamefont {Bailey}\ \emph {et~al.}(2013)\citenamefont {Bailey},
  \citenamefont {Everett},\ and\ \citenamefont
  {Overduin}}]{PhysRevD.88.102001}%
  \BibitemOpen
  \bibfield  {author} {\bibinfo {author} {\bibfnamefont {Q.~G.}\ \bibnamefont
  {Bailey}}, \bibinfo {author} {\bibfnamefont {R.~D.}\ \bibnamefont {Everett}},
  \ and\ \bibinfo {author} {\bibfnamefont {J.~M.}\ \bibnamefont {Overduin}},\
  }\href {\doibase 10.1103/PhysRevD.88.102001} {\bibfield  {journal} {\bibinfo
  {journal} {Phys. Rev. D}\ }\textbf {\bibinfo {volume} {88}},\ \bibinfo
  {pages} {102001} (\bibinfo {year} {2013})}\BibitemShut {NoStop}%
\bibitem [{\citenamefont {Shao}(2014)}]{PhysRevLett.112.111103}%
  \BibitemOpen
  \bibfield  {author} {\bibinfo {author} {\bibfnamefont {L.}~\bibnamefont
  {Shao}},\ }\href {\doibase 10.1103/PhysRevLett.112.111103} {\bibfield
  {journal} {\bibinfo  {journal} {Phys. Rev. Lett.}\ }\textbf {\bibinfo
  {volume} {112}},\ \bibinfo {pages} {111103} (\bibinfo {year}
  {2014})}\BibitemShut {NoStop}%
\bibitem [{\citenamefont {Kosteleck{\'y}}\ and\ \citenamefont
  {Tasson}(2015)}]{kostelecky2015constraints}%
  \BibitemOpen
  \bibfield  {author} {\bibinfo {author} {\bibfnamefont {V.~A.}\ \bibnamefont
  {Kosteleck{\'y}}}\ and\ \bibinfo {author} {\bibfnamefont {J.~D.}\
  \bibnamefont {Tasson}},\ }\href {\doibase
  https://doi.org/10.1016/j.physletb.2015.08.060} {\bibfield  {journal}
  {\bibinfo  {journal} {Phys. Lett. B}\ }\textbf {\bibinfo {volume} {749}},\
  \bibinfo {pages} {551} (\bibinfo {year} {2015})}\BibitemShut {NoStop}%
\bibitem [{\citenamefont {Yunes}\ \emph {et~al.}(2016)\citenamefont {Yunes},
  \citenamefont {Yagi},\ and\ \citenamefont {Pretorius}}]{PhysRevD.94.084002}%
  \BibitemOpen
  \bibfield  {author} {\bibinfo {author} {\bibfnamefont {N.}~\bibnamefont
  {Yunes}}, \bibinfo {author} {\bibfnamefont {K.}~\bibnamefont {Yagi}}, \ and\
  \bibinfo {author} {\bibfnamefont {F.}~\bibnamefont {Pretorius}},\ }\href
  {\doibase 10.1103/PhysRevD.94.084002} {\bibfield  {journal} {\bibinfo
  {journal} {Phys. Rev. D}\ }\textbf {\bibinfo {volume} {94}},\ \bibinfo
  {pages} {084002} (\bibinfo {year} {2016})}\BibitemShut {NoStop}%
\bibitem [{\citenamefont {Shao}\ \emph {et~al.}(2016)\citenamefont {Shao} \emph
  {et~al.}}]{PhysRevLett.117.071102}%
  \BibitemOpen
  \bibfield  {author} {\bibinfo {author} {\bibfnamefont {C.-G.}\ \bibnamefont
  {Shao}} \emph {et~al.},\ }\href {\doibase 10.1103/PhysRevLett.117.071102}
  {\bibfield  {journal} {\bibinfo  {journal} {Phys. Rev. Lett.}\ }\textbf
  {\bibinfo {volume} {117}},\ \bibinfo {pages} {071102} (\bibinfo {year}
  {2016})}\BibitemShut {NoStop}%
\bibitem [{\citenamefont {Flowers}\ \emph {et~al.}(2017)\citenamefont
  {Flowers}, \citenamefont {Goodge},\ and\ \citenamefont
  {Tasson}}]{PhysRevLett.119.201101}%
  \BibitemOpen
  \bibfield  {author} {\bibinfo {author} {\bibfnamefont {N.~A.}\ \bibnamefont
  {Flowers}}, \bibinfo {author} {\bibfnamefont {C.}~\bibnamefont {Goodge}}, \
  and\ \bibinfo {author} {\bibfnamefont {J.~D.}\ \bibnamefont {Tasson}},\
  }\href {\doibase 10.1103/PhysRevLett.119.201101} {\bibfield  {journal}
  {\bibinfo  {journal} {Phys. Rev. Lett.}\ }\textbf {\bibinfo {volume} {119}},\
  \bibinfo {pages} {201101} (\bibinfo {year} {2017})}\BibitemShut {NoStop}%
\bibitem [{\citenamefont {Bourgoin}\ \emph {et~al.}(2017)\citenamefont
  {Bourgoin} \emph {et~al.}}]{PhysRevLett.119.201102}%
  \BibitemOpen
  \bibfield  {author} {\bibinfo {author} {\bibfnamefont {A.}~\bibnamefont
  {Bourgoin}} \emph {et~al.},\ }\href {\doibase 10.1103/PhysRevLett.119.201102}
  {\bibfield  {journal} {\bibinfo  {journal} {Phys. Rev. Lett.}\ }\textbf
  {\bibinfo {volume} {119}},\ \bibinfo {pages} {201102} (\bibinfo {year}
  {2017})}\BibitemShut {NoStop}%
\bibitem [{\citenamefont {Shao}\ \emph {et~al.}(2018)\citenamefont {Shao} \emph
  {et~al.}}]{PhysRevD.97.024019}%
  \BibitemOpen
  \bibfield  {author} {\bibinfo {author} {\bibfnamefont {C.-G.}\ \bibnamefont
  {Shao}} \emph {et~al.},\ }\href {\doibase 10.1103/PhysRevD.97.024019}
  {\bibfield  {journal} {\bibinfo  {journal} {Phys. Rev. D}\ }\textbf {\bibinfo
  {volume} {97}},\ \bibinfo {pages} {024019} (\bibinfo {year}
  {2018})}\BibitemShut {NoStop}%
\bibitem [{\citenamefont {Mewes}(2019)}]{PhysRevD.99.104062}%
  \BibitemOpen
  \bibfield  {author} {\bibinfo {author} {\bibfnamefont {M.}~\bibnamefont
  {Mewes}},\ }\href {\doibase 10.1103/PhysRevD.99.104062} {\bibfield  {journal}
  {\bibinfo  {journal} {Phys. Rev. D}\ }\textbf {\bibinfo {volume} {99}},\
  \bibinfo {pages} {104062} (\bibinfo {year} {2019})}\BibitemShut {NoStop}%
\bibitem [{\citenamefont {Escobar}\ and\ \citenamefont
  {Mart\'{\i}n-Ruiz}(2019)}]{PhysRevD.99.075032}%
  \BibitemOpen
  \bibfield  {author} {\bibinfo {author} {\bibfnamefont {C.~A.}\ \bibnamefont
  {Escobar}}\ and\ \bibinfo {author} {\bibfnamefont {A.}~\bibnamefont
  {Mart\'{\i}n-Ruiz}},\ }\href {\doibase 10.1103/PhysRevD.99.075032} {\bibfield
   {journal} {\bibinfo  {journal} {Phys. Rev. D}\ }\textbf {\bibinfo {volume}
  {99}},\ \bibinfo {pages} {075032} (\bibinfo {year} {2019})}\BibitemShut
  {NoStop}%
\bibitem [{\citenamefont {Hees}\ \emph {et~al.}(2016)\citenamefont {Hees} \emph
  {et~al.}}]{reviewbounds}%
  \BibitemOpen
  \bibfield  {author} {\bibinfo {author} {\bibfnamefont {A.}~\bibnamefont
  {Hees}} \emph {et~al.},\ }\href {\doibase 10.3390/universe2040030} {\bibfield
   {journal} {\bibinfo  {journal} {Universe}\ }\textbf {\bibinfo {volume}
  {2}},\ \bibinfo {pages} {30} (\bibinfo {year} {2016})}\BibitemShut {NoStop}%
\bibitem [{\citenamefont {Bonder}(2015)}]{Bonder2015}%
  \BibitemOpen
  \bibfield  {author} {\bibinfo {author} {\bibfnamefont {Y.}~\bibnamefont
  {Bonder}},\ }\href {\doibase 10.1103/PhysRevD.91.125002} {\bibfield
  {journal} {\bibinfo  {journal} {Phys. Rev. D}\ }\textbf {\bibinfo {volume}
  {91}},\ \bibinfo {pages} {125002} (\bibinfo {year} {2015})}\BibitemShut
  {NoStop}%
\bibitem [{\citenamefont {Corral}\ and\ \citenamefont
  {Bonder}(2019)}]{Cristobal1}%
  \BibitemOpen
  \bibfield  {author} {\bibinfo {author} {\bibfnamefont {C.}~\bibnamefont
  {Corral}}\ and\ \bibinfo {author} {\bibfnamefont {Y.}~\bibnamefont
  {Bonder}},\ }\href {\doibase 10.1088/1361-6382/aafce1} {\bibfield  {journal}
  {\bibinfo  {journal} {Class. Quantum Grav.}\ }\textbf {\bibinfo {volume}
  {36}},\ \bibinfo {pages} {045002} (\bibinfo {year} {2019})}\BibitemShut
  {NoStop}%
\bibitem [{\citenamefont {Bluhm}(2015)}]{PhysRevD.91.065034}%
  \BibitemOpen
  \bibfield  {author} {\bibinfo {author} {\bibfnamefont {R.}~\bibnamefont
  {Bluhm}},\ }\href {\doibase 10.1103/PhysRevD.91.065034} {\bibfield  {journal}
  {\bibinfo  {journal} {Phys. Rev. D}\ }\textbf {\bibinfo {volume} {91}},\
  \bibinfo {pages} {065034} (\bibinfo {year} {2015})}\BibitemShut {NoStop}%
\bibitem [{\citenamefont {Bluhm}\ and\ \citenamefont
  {Kosteleck\'y}(2005)}]{KosteleckyBlumm1}%
  \BibitemOpen
  \bibfield  {author} {\bibinfo {author} {\bibfnamefont {R.}~\bibnamefont
  {Bluhm}}\ and\ \bibinfo {author} {\bibfnamefont {V.~A.}\ \bibnamefont
  {Kosteleck\'y}},\ }\href {\doibase 10.1103/PhysRevD.71.065008} {\bibfield
  {journal} {\bibinfo  {journal} {Phys. Rev. D}\ }\textbf {\bibinfo {volume}
  {71}},\ \bibinfo {pages} {065008} (\bibinfo {year} {2005})}\BibitemShut
  {NoStop}%
\bibitem [{\citenamefont {Bluhm}\ \emph
  {et~al.}(2008{\natexlab{a}})\citenamefont {Bluhm}, \citenamefont {Fung},\
  and\ \citenamefont {Kosteleck\'y}}]{KosteleckyBlumm2}%
  \BibitemOpen
  \bibfield  {author} {\bibinfo {author} {\bibfnamefont {R.}~\bibnamefont
  {Bluhm}}, \bibinfo {author} {\bibfnamefont {S.-H.}\ \bibnamefont {Fung}}, \
  and\ \bibinfo {author} {\bibfnamefont {V.~A.}\ \bibnamefont {Kosteleck\'y}},\
  }\href {\doibase 10.1103/PhysRevD.77.065020} {\bibfield  {journal} {\bibinfo
  {journal} {Phys. Rev. D}\ }\textbf {\bibinfo {volume} {77}},\ \bibinfo
  {pages} {065020} (\bibinfo {year} {2008}{\natexlab{a}})}\BibitemShut
  {NoStop}%
\bibitem [{\citenamefont {Bluhm}\ \emph
  {et~al.}(2008{\natexlab{b}})\citenamefont {Bluhm}, \citenamefont {Gagne},
  \citenamefont {Potting},\ and\ \citenamefont
  {Vrublevskis}}]{PhysRevD.77.125007}%
  \BibitemOpen
  \bibfield  {author} {\bibinfo {author} {\bibfnamefont {R.}~\bibnamefont
  {Bluhm}}, \bibinfo {author} {\bibfnamefont {N.~L.}\ \bibnamefont {Gagne}},
  \bibinfo {author} {\bibfnamefont {R.}~\bibnamefont {Potting}}, \ and\
  \bibinfo {author} {\bibfnamefont {A.}~\bibnamefont {Vrublevskis}},\ }\href
  {\doibase 10.1103/PhysRevD.77.125007} {\bibfield  {journal} {\bibinfo
  {journal} {Phys. Rev. D}\ }\textbf {\bibinfo {volume} {77}},\ \bibinfo
  {pages} {125007} (\bibinfo {year} {2008}{\natexlab{b}})}\BibitemShut
  {NoStop}%
\bibitem [{\citenamefont {Bluhm}\ \emph {et~al.}(2009)\citenamefont {Bluhm},
  \citenamefont {Gagne}, \citenamefont {Potting},\ and\ \citenamefont
  {Vrublevskis}}]{PhysRevD.79.029902}%
  \BibitemOpen
  \bibfield  {author} {\bibinfo {author} {\bibfnamefont {R.}~\bibnamefont
  {Bluhm}}, \bibinfo {author} {\bibfnamefont {N.~L.}\ \bibnamefont {Gagne}},
  \bibinfo {author} {\bibfnamefont {R.}~\bibnamefont {Potting}}, \ and\
  \bibinfo {author} {\bibfnamefont {A.}~\bibnamefont {Vrublevskis}},\ }\href
  {\doibase 10.1103/PhysRevD.79.029902} {\bibfield  {journal} {\bibinfo
  {journal} {Phys. Rev. D}\ }\textbf {\bibinfo {volume} {79}},\ \bibinfo
  {pages} {029902} (\bibinfo {year} {2009})}\BibitemShut {NoStop}%
\bibitem [{\citenamefont {Bonder}\ and\ \citenamefont
  {Escobar}(2016)}]{BonderEscobar}%
  \BibitemOpen
  \bibfield  {author} {\bibinfo {author} {\bibfnamefont {Y.}~\bibnamefont
  {Bonder}}\ and\ \bibinfo {author} {\bibfnamefont {C.~A.}\ \bibnamefont
  {Escobar}},\ }\href {\doibase 10.1103/PhysRevD.93.025020} {\bibfield
  {journal} {\bibinfo  {journal} {Phys. Rev. D}\ }\textbf {\bibinfo {volume}
  {93}},\ \bibinfo {pages} {025020} (\bibinfo {year} {2016})}\BibitemShut
  {NoStop}%
\bibitem [{\citenamefont {Bailey}\ and\ \citenamefont
  {Kosteleck\'y}(2006)}]{BaileyKostelecky2006}%
  \BibitemOpen
  \bibfield  {author} {\bibinfo {author} {\bibfnamefont {Q.~G.}\ \bibnamefont
  {Bailey}}\ and\ \bibinfo {author} {\bibfnamefont {V.~A.}\ \bibnamefont
  {Kosteleck\'y}},\ }\href {\doibase 10.1103/PhysRevD.74.045001} {\bibfield
  {journal} {\bibinfo  {journal} {Phys. Rev. D}\ }\textbf {\bibinfo {volume}
  {74}},\ \bibinfo {pages} {045001} (\bibinfo {year} {2006})}\BibitemShut
  {NoStop}%
\bibitem [{\citenamefont {Bailey}(2009)}]{tpuzz2}%
  \BibitemOpen
  \bibfield  {author} {\bibinfo {author} {\bibfnamefont {Q.~G.}\ \bibnamefont
  {Bailey}},\ }\href {\doibase 10.1103/PhysRevD.80.044004} {\bibfield
  {journal} {\bibinfo  {journal} {Phys. Rev. D}\ }\textbf {\bibinfo {volume}
  {80}},\ \bibinfo {pages} {044004} (\bibinfo {year} {2009})}\BibitemShut
  {NoStop}%
\bibitem [{\citenamefont {Altschul}\ \emph {et~al.}(2010)\citenamefont
  {Altschul}, \citenamefont {Bailey},\ and\ \citenamefont
  {Kosteleck\'y}}]{AltschulBaileyKostelecky2010}%
  \BibitemOpen
  \bibfield  {author} {\bibinfo {author} {\bibfnamefont {B.}~\bibnamefont
  {Altschul}}, \bibinfo {author} {\bibfnamefont {Q.~G.}\ \bibnamefont
  {Bailey}}, \ and\ \bibinfo {author} {\bibfnamefont {V.~A.}\ \bibnamefont
  {Kosteleck\'y}},\ }\href {\doibase 10.1103/PhysRevD.81.065028} {\bibfield
  {journal} {\bibinfo  {journal} {Phys. Rev. D}\ }\textbf {\bibinfo {volume}
  {81}},\ \bibinfo {pages} {065028} (\bibinfo {year} {2010})}\BibitemShut
  {NoStop}%
\bibitem [{\citenamefont {Bailey}(2010)}]{tpuzz3}%
  \BibitemOpen
  \bibfield  {author} {\bibinfo {author} {\bibfnamefont {Q.~G.}\ \bibnamefont
  {Bailey}},\ }\href {\doibase 10.1103/PhysRevD.82.065012} {\bibfield
  {journal} {\bibinfo  {journal} {Phys. Rev. D}\ }\textbf {\bibinfo {volume}
  {82}},\ \bibinfo {pages} {065012} (\bibinfo {year} {2010})}\BibitemShut
  {NoStop}%
\bibitem [{\citenamefont {Tso}\ and\ \citenamefont {Bailey}(2011)}]{tpuzz4}%
  \BibitemOpen
  \bibfield  {author} {\bibinfo {author} {\bibfnamefont {R.}~\bibnamefont
  {Tso}}\ and\ \bibinfo {author} {\bibfnamefont {Q.~G.}\ \bibnamefont
  {Bailey}},\ }\href {\doibase 10.1103/PhysRevD.84.085025} {\bibfield
  {journal} {\bibinfo  {journal} {Phys. Rev. D}\ }\textbf {\bibinfo {volume}
  {84}},\ \bibinfo {pages} {085025} (\bibinfo {year} {2011})}\BibitemShut
  {NoStop}%
\bibitem [{\citenamefont {Tasson}(2012)}]{tpuzz5}%
  \BibitemOpen
  \bibfield  {author} {\bibinfo {author} {\bibfnamefont {J.~D.}\ \bibnamefont
  {Tasson}},\ }\href {\doibase 10.1103/PhysRevD.86.124021} {\bibfield
  {journal} {\bibinfo  {journal} {Phys. Rev. D}\ }\textbf {\bibinfo {volume}
  {86}},\ \bibinfo {pages} {124021} (\bibinfo {year} {2012})}\BibitemShut
  {NoStop}%
\bibitem [{\citenamefont {Bailey}\ \emph {et~al.}(2015)\citenamefont {Bailey},
  \citenamefont {Kosteleck\'y},\ and\ \citenamefont
  {Xu}}]{BaileyKosteleckyXu2015}%
  \BibitemOpen
  \bibfield  {author} {\bibinfo {author} {\bibfnamefont {Q.~G.}\ \bibnamefont
  {Bailey}}, \bibinfo {author} {\bibfnamefont {V.~A.}\ \bibnamefont
  {Kosteleck\'y}}, \ and\ \bibinfo {author} {\bibfnamefont {R.}~\bibnamefont
  {Xu}},\ }\href {\doibase 10.1103/PhysRevD.91.022006} {\bibfield  {journal}
  {\bibinfo  {journal} {Phys. Rev. D}\ }\textbf {\bibinfo {volume} {91}},\
  \bibinfo {pages} {022006} (\bibinfo {year} {2015})}\BibitemShut {NoStop}%
\bibitem [{\citenamefont {Bonder}\ and\ \citenamefont
  {Le\'on}(2017)}]{Gabriel}%
  \BibitemOpen
  \bibfield  {author} {\bibinfo {author} {\bibfnamefont {Y.}~\bibnamefont
  {Bonder}}\ and\ \bibinfo {author} {\bibfnamefont {G.}~\bibnamefont
  {Le\'on}},\ }\href {\doibase 10.1103/PhysRevD.96.044036} {\bibfield
  {journal} {\bibinfo  {journal} {Phys. Rev. D}\ }\textbf {\bibinfo {volume}
  {96}},\ \bibinfo {pages} {044036} (\bibinfo {year} {2017})}\BibitemShut
  {NoStop}%
\bibitem [{\citenamefont {Bluhm}\ \emph {et~al.}(2019)\citenamefont {Bluhm},
  \citenamefont {Bossi},\ and\ \citenamefont {Wen}}]{Explicit}%
  \BibitemOpen
  \bibfield  {author} {\bibinfo {author} {\bibfnamefont {R.}~\bibnamefont
  {Bluhm}}, \bibinfo {author} {\bibfnamefont {H.}~\bibnamefont {Bossi}}, \ and\
  \bibinfo {author} {\bibfnamefont {Y.}~\bibnamefont {Wen}},\ }\href {\doibase
  10.1103/PhysRevD.100.084022} {\bibfield  {journal} {\bibinfo  {journal}
  {Phys. Rev. D}\ }\textbf {\bibinfo {volume} {100}},\ \bibinfo {pages}
  {084022} (\bibinfo {year} {2019})}\BibitemShut {NoStop}%
\bibitem [{\citenamefont {Bonder}\ and\ \citenamefont
  {Corral}(2018)}]{Cristobal2}%
  \BibitemOpen
  \bibfield  {author} {\bibinfo {author} {\bibfnamefont {Y.}~\bibnamefont
  {Bonder}}\ and\ \bibinfo {author} {\bibfnamefont {C.}~\bibnamefont
  {Corral}},\ }\href {https://www.mdpi.com/2073-8994/10/10/433} {\bibfield
  {journal} {\bibinfo  {journal} {Symmetry}\ }\textbf {\bibinfo {volume}
  {10}},\ \bibinfo {pages} {433} (\bibinfo {year} {2018})}\BibitemShut
  {NoStop}%
\bibitem [{\citenamefont {Frolov}\ and\ \citenamefont
  {Zelnikov}(2011)}]{Frolov}%
  \BibitemOpen
  \bibfield  {author} {\bibinfo {author} {\bibfnamefont {V.~P.}\ \bibnamefont
  {Frolov}}\ and\ \bibinfo {author} {\bibfnamefont {A.}~\bibnamefont
  {Zelnikov}},\ }\href@noop {} {\emph {\bibinfo {title} {{Introduction to Black
  Hole Physics}}}}\ (\bibinfo  {publisher} {OUP Oxford},\ \bibinfo {year}
  {2011})\BibitemShut {NoStop}%
\bibitem [{\citenamefont {Hall}(2004)}]{Hall}%
  \BibitemOpen
  \bibfield  {author} {\bibinfo {author} {\bibfnamefont {G.~S.}\ \bibnamefont
  {Hall}},\ }\href@noop {} {\emph {\bibinfo {title} {Symmetries and Curvature
  Structure in General Relativity}}}\ (\bibinfo  {publisher} {World
  Scientific},\ \bibinfo {year} {2004})\BibitemShut {NoStop}%
\bibitem [{\citenamefont {Bonder}\ and\ \citenamefont
  {Sudarsky}(2008)}]{Bonder2008}%
  \BibitemOpen
  \bibfield  {author} {\bibinfo {author} {\bibfnamefont {Y.}~\bibnamefont
  {Bonder}}\ and\ \bibinfo {author} {\bibfnamefont {D.}~\bibnamefont
  {Sudarsky}},\ }\href {\doibase 10.1088/0264-9381/25/10/105017} {\bibfield
  {journal} {\bibinfo  {journal} {Class. Quantum Grav.}\ }\textbf {\bibinfo
  {volume} {25}},\ \bibinfo {pages} {105017} (\bibinfo {year}
  {2008})}\BibitemShut {NoStop}%
\bibitem [{\citenamefont {Salgado}(2003)}]{Marcelo}%
  \BibitemOpen
  \bibfield  {author} {\bibinfo {author} {\bibfnamefont {M.}~\bibnamefont
  {Salgado}},\ }\href {\doibase 10.1088/0264-9381/20/21/003} {\bibfield
  {journal} {\bibinfo  {journal} {Class. Quantum Grav.}\ }\textbf {\bibinfo
  {volume} {20}},\ \bibinfo {pages} {4551} (\bibinfo {year}
  {2003})}\BibitemShut {NoStop}%
\bibitem [{\citenamefont {Jacobson}(2007)}]{Jacobson}%
  \BibitemOpen
  \bibfield  {author} {\bibinfo {author} {\bibfnamefont {T.}~\bibnamefont
  {Jacobson}},\ }\href {\doibase 10.1088/0264-9381/24/22/n02} {\bibfield
  {journal} {\bibinfo  {journal} {Class. Quantum Grav.}\ }\textbf {\bibinfo
  {volume} {24}},\ \bibinfo {pages} {5717} (\bibinfo {year}
  {2007})}\BibitemShut {NoStop}%
\bibitem [{\citenamefont {Will}(2018)}]{Will}%
  \BibitemOpen
  \bibfield  {author} {\bibinfo {author} {\bibfnamefont {C.~M.}\ \bibnamefont
  {Will}},\ }\href@noop {} {\emph {\bibinfo {title} {{Theory and Experiment in
  Gravitational Physics}}}}\ (\bibinfo  {publisher} {Cambridge University
  Press},\ \bibinfo {year} {2018})\BibitemShut {NoStop}%
\bibitem [{\citenamefont {Hartle}(2003)}]{Hartle}%
  \BibitemOpen
  \bibfield  {author} {\bibinfo {author} {\bibfnamefont {J.~B.}\ \bibnamefont
  {Hartle}},\ }\href@noop {} {\emph {\bibinfo {title} {{Gravity: An
  Introduction to Einstein's General Relativity}}}}\ (\bibinfo  {publisher}
  {Addison Wesley},\ \bibinfo {year} {2003})\BibitemShut {NoStop}%
\bibitem [{\citenamefont {Peterson}\ and\ \citenamefont
  {Bonder}(2020)}]{TKilling}%
  \BibitemOpen
  \bibfield  {author} {\bibinfo {author} {\bibfnamefont {C.}~\bibnamefont
  {Peterson}}\ and\ \bibinfo {author} {\bibfnamefont {Y.}~\bibnamefont
  {Bonder}},\ }\href {\doibase 10.1142/S0217732320500522} {\bibfield  {journal}
  {\bibinfo  {journal} {Mod. Phys. Lett. A}\ }\textbf {\bibinfo {volume}
  {33}},\ \bibinfo {pages} {2050052} (\bibinfo {year} {2020})}\BibitemShut
  {NoStop}%
\bibitem [{\citenamefont {Delva}\ \emph {et~al.}(2018)\citenamefont {Delva}
  \emph {et~al.}}]{GREAT}%
  \BibitemOpen
  \bibfield  {author} {\bibinfo {author} {\bibfnamefont {P.}~\bibnamefont
  {Delva}} \emph {et~al.},\ }\href {\doibase 10.1103/PhysRevLett.121.231101}
  {\bibfield  {journal} {\bibinfo  {journal} {Phys. Rev. Lett.}\ }\textbf
  {\bibinfo {volume} {121}},\ \bibinfo {pages} {231101} (\bibinfo {year}
  {2018})}\BibitemShut {NoStop}%
\bibitem [{\citenamefont {Carlier}(2016)}]{DatosGalileo}%
  \BibitemOpen
  \bibfield  {author} {\bibinfo {author} {\bibfnamefont {N.}~\bibnamefont
  {Carlier}},\ }in\ \href@noop {} {\emph {\bibinfo {booktitle} {14th
  International Conference on Space Operations}}}\ (\bibinfo {year} {2016})\
  p.\ \bibinfo {pages} {2561}\BibitemShut {NoStop}%
\bibitem [{\citenamefont {Buchdahl}(1985)}]{Buchdahl}%
  \BibitemOpen
  \bibfield  {author} {\bibinfo {author} {\bibfnamefont {H.}~\bibnamefont
  {Buchdahl}},\ }\href {https://doi.org/10.1007/BF00670880} {\bibfield
  {journal} {\bibinfo  {journal} {Int. J. Theo. Phys.}\ }\textbf {\bibinfo
  {volume} {24}},\ \bibinfo {pages} {731} (\bibinfo {year} {1985})}\BibitemShut
  {NoStop}%
\bibitem [{\citenamefont {Bertotti}\ \emph {et~al.}(2003)\citenamefont
  {Bertotti}, \citenamefont {Iess},\ and\ \citenamefont {Tortora}}]{Shapiro1}%
  \BibitemOpen
  \bibfield  {author} {\bibinfo {author} {\bibfnamefont {B.}~\bibnamefont
  {Bertotti}}, \bibinfo {author} {\bibfnamefont {L.}~\bibnamefont {Iess}}, \
  and\ \bibinfo {author} {\bibfnamefont {P.}~\bibnamefont {Tortora}},\ }\href
  {https://doi.org/10.1038/nature01997} {\bibfield  {journal} {\bibinfo
  {journal} {Nature}\ }\textbf {\bibinfo {volume} {425}},\ \bibinfo {pages}
  {374} (\bibinfo {year} {2003})}\BibitemShut {NoStop}%
\bibitem [{\citenamefont {Imperi}\ and\ \citenamefont {Iess}(2017)}]{Shapiro2}%
  \BibitemOpen
  \bibfield  {author} {\bibinfo {author} {\bibfnamefont {L.}~\bibnamefont
  {Imperi}}\ and\ \bibinfo {author} {\bibfnamefont {L.}~\bibnamefont {Iess}},\
  }\href {\doibase 10.1088/1361-6382/aa606d} {\bibfield  {journal} {\bibinfo
  {journal} {Class. Quantum Grav.}\ }\textbf {\bibinfo {volume} {34}},\
  \bibinfo {pages} {075002} (\bibinfo {year} {2017})}\BibitemShut {NoStop}%
\end{thebibliography}%

\end{document}